%% file: main.tex
\documentclass[acmlarge]{acmart}
\include{Preamble}

\usepackage{siunitx}
\usepackage{upgreek}

\begin{document}

\title{End-to-End Simultaneous Learning of Single-particle Orientation and 3D Map Reconstruction from Cryo-electron Microscopy Data}

\author{Youssef S. G. Nashed}
\authornote{Both authors contributed equally to this research.}
\email{ynashed@slac.stanford.edu}
\orcid{0000-0001-6146-3939}
\affiliation{%
  \institution{Machine Learning Initiative, SLAC National Accelerator Laboratory}
  \city{Menlo Park}
  \state{California}
  \country{USA}
}

\author{Frederic Poitevin}
\authornotemark[1]
\email{frederic.poitevin@stanford.edu}
\orcid{0000-0002-3181-8652}
\affiliation{%
  \institution{Department of LCLS Data Analytics, SLAC National Accelerator Laboratory}
  \city{Menlo Park}
  \state{California}
  \country{USA}
}

\author{Harshit Gupta}
\affiliation{%
  \institution{Machine Learning Initiative, SLAC National Accelerator Laboratory}
  \city{Menlo Park}
  \state{California}
  \country{USA}
}

\author{Geoffrey Woollard}
\affiliation{%
  \institution{Department of Mathematics, University of British Columbia}
  \city{Vancouver}
  \country{Canada}
}

\author{Michael Kagan}
\affiliation{%
  \institution{Fundamental Physics Directorate, SLAC National Accelerator Laboratory}
  \city{Menlo Park}
  \state{California}
  \country{USA}
}

\author{Chuck Yoon}
\affiliation{%
  \institution{Department of LCLS Data Analytics, SLAC National Accelerator Laboratory}
  \city{Menlo Park}
  \state{California}
  \country{USA}
}

\author{Daniel Ratner}
\authornote{Corresponding author.}
\affiliation{%
  \institution{Machine Learning Initiative, SLAC National Accelerator Laboratory}
  \city{Menlo Park}
  \state{California}
  \country{USA}
}


\begin{abstract}
Cryogenic electron microscopy (cryo-EM) provides images from different copies of the same biomolecule in arbitrary orientations.  Here, we present an end-to-end unsupervised approach that learns individual particle orientations from cryo-EM data while reconstructing the average 3D map of the biomolecule, starting from a random initialization. The approach relies on an auto-encoder architecture where the latent space is explicitly interpreted as orientations used by the decoder to form an image according to the linear projection model. We evaluate our method on simulated data and show that it is able to reconstruct 3D particle maps from noisy- and CTF-corrupted 2D projection images of unknown particle orientations.
\end{abstract}

\maketitle
\section{Introduction}

Determining the structure of biomolecules is an important step towards understanding their functional mechanism or designing new drugs. 
Structural determination techniques such as nuclear magnetic resonance spectroscopy, X-ray diffraction (XRD) crystallography or cryogenic electron microscopy (cryo-EM) have been very successful over the years \cite{noauthor_celebration_2021}. Of those techniques, cryo-EM has been used increasingly in recent years to determine structures of many biomolecules at near-atomic resolution. This revolution has been possible because of better hardware and software techniques \cite{kuhlbrandt_resolution_2014}. Cryo-EM is the fastest growing technique in terms of structures deposited in the Protein Data Bank, projected to be comparable to XRD within a few years. While the typical resolution reported for cryo-EM structures is still worse than those reported for XRD structures, recent developments have proven possible to distinguish individual atoms in cryo-EM maps \cite{nakane_single-particle_2020}. Whereas XRD simultaneously measures millions of copies of a molecule, cryo-EM is a single-particle imaging (SPI) technique, with each cryo-EM image corresponding to a single molecular copy. While SPI reconstruction methods must learn the conformation and orientation of individual particles, this challenge is also an opportunity: rather than learning an average structure, SPI methods can resolve biologically relevant dynamics and avoid blurring from averaging over heterogeneous particles \cite{ourmazd_cryo-em_2019}. In this paper, we discuss a new approach to solving the single-particle orientation problem using a neural-network auto-encoder.

\subsection{Cryo-EM 3D Reconstruction}

In cryo-EM, a large number ($10^4-10^7$) of identical copies of the same biomolecule are first frozen in a thin vitreous layer of ice. A beam of electron then goes through this sample and is acquired by a detector. It is typically assumed that only the phase of the incoming beam has been changed by the electrostatic potential of the sample, not its amplitude. Since the sample is typically prepared so the material that the electrons go through is very thin, it is also usually assumed that the linear projection approximation can be used to model the image formation process. The resulting image is called a micrograph. Individual particles are located and extracted from the micrograph as square patches of identical size $N\times N$ resulting in a particle stack which constitutes the starting dataset for tomographic reconstruction of the particle $N\times N\times N$ volume. Tomographic reconstruction is an ill-posed problem that faces multiple challenges. The individual pose (orientation and translational shift) of each particle in the stack is not known and needs to be estimated from images that have been corrupted in two different ways. First, the image is convolved with the point-spread function (PSF) of the microscope which acts as a filter reducing the information content of each image across its spectrum. Second, the main part of the signal in the image comes from the surrounding ice which is a major source of noise, in addition to other sources of noise such as shot noise resulting from dose fractionation and detector response. Indeed a balance between radiation damage and getting signal at all needs to be found during data collection. These difficulties make cryo-EM 3D reconstruction a very challenging inverse problem.

\subsection{Related work}

Many methods have been developed to tackle cryo-EM tomographic reconstruction \cite{sorzano_xmipp_2004, grigorieff_frealign_2007, shaikh_spider_2008, tang_eman2_2007, hohn_sparx_2007, scheres_relion_2012, de_la_rosa-trevin_scipion_2016, punjani_cryosparc_2017}, including Bayesian formulations \cite{dempster_maximum_1977, sigworth_maximum-likelihood_1998, scheres_relion_2012, punjani_cryosparc_2017}, projection-matching strategies \cite{penczek_ribosome_1994, baker_model-based_1996}, common-lines approaches \cite{vainshtein_determination_1986} or other techniques \cite{mallick_structure_2006, singer_detecting_2010, wang_orientation_2013, greenberg_common_2017, pragier_common_2019, zehni_joint_2020}. The Bayesian approach has been popularized by the widely used software RELION \cite{scheres_bayesian_2012} which performs maximum-a-posteriori (MAP) optimization through Expectation-Maximization (EM) to reconstruct a map which maximizes the likelihood of the acquired data while still meeting some \emph{a priori} condition about the map. In EM, during the Expectation-step, a conditional distribution over the poses and shifts is estimated for each projection using the current estimate of the map. In the next Maximization-step, these estimated distributions are used to update the map. These two steps are performed iteratively until meeting a convergence criterion. MAP optimization does not guarantee global convergence and therefore its reconstruction is dependent on the quality of the initial map.  First implementations of the method suffered from their lack of robustness to map initialization, so a competing software cryoSPARC \cite{punjani_cryosparc_2017} proposed optimizing a MAP solution \emph{ab initio} through stochastic gradient descent (SGD) - however, in doing so, individual poses are not estimated explicitly, thereby limiting the achievable resolution. The resulting low-resolution map can then be further refined with the EM algorithm. Because the EM reconstruction approach requires estimation of conditional distribution on poses for each measurement, the number of variables to determine grows directly with the number of measurements. 

Neural networks (NNs) with their state-of-the-art performance on various different inverse problems \cite{litjens_survey_2017, mccann_convolutional_2017} have recently been introduced in the cryo-EM processing pipeline, mainly in pre-processing steps such as denoising of the micrograph \cite{bepler_topaz-denoise_2020} and particle picking \cite{wang_deeppicker_2016, zhu_deep_2017, tegunov_real-time_2019, wagner_sphire-cryolo_2019, bepler_positive-unlabeled_2019}.  
More recently a few methods using NNs have been proposed to solve the cryo-EM reconstruction problem \cite{zhong_cryodrgn_2021, rosenbaum_inferring_2021, gupta_cryogan_2020, gupta_multi-cryogan_2020, punjani_3d_2021, miolane_estimation_2021}. These methods require no prior training and are fully unsupervised. In \cite{zhong_cryodrgn_2021} a modified variational autoencoder (VAE) \cite{kingma_auto-encoding_2014} is used to reconstruct continuous conformations of dynamic biomolecules. In \cite{gupta_cryogan_2020, gupta_multi-cryogan_2020}, generative adversarial networks (GANs) \cite{goodfellow_generative_2014} are modified to reconstruct structure of biomolecules with single and continuous conformations, respectively. In \cite{miolane_estimation_2021} a composition of GANs and VAE is used to find the latent variables that explain the data, which then can be further used for reconstruction. In contrast to these methods, \cite{punjani_3d_2021} argues that amortized inference as done by VAEs does not yield results that are precise enough and instead perform direct inference of the conformational coordinate using an auto-decoder approach. None of these methods use NNs to estimate the poses for each projections. Instead they either use an external traditional pose estimation routine \cite{zhong_cryodrgn_2021} or bypass the pose assignment step altogether \cite{gupta_cryogan_2020, gupta_multi-cryogan_2020}. 

\subsection{Contributions}

In this paper, we propose a method which uses a modified auto-encoder to simultaneously reconstruct the 3D map and estimate individual poses. Our method is fully unsupervised, and does not require any prior training nor an \textit{ab-initio} solution.
We use a combination of an encoder parameterized by an NN and a cryo-EM physics based decoder parameterized by a learnable 3D structure. The encoder outputs the estimated pose for a given input measurement, which is then fed to the decoder that outputs the simulated cryo-EM projection from the current structure. The encodings are constrained to correspond to the pose by the physics-based decoder. The weights of the encoder and the 3D structure in the decoder are simultaneously optimized in order to decrease the error between the given projection (input of the encoder) and the output of the decoder.

We leverage this auto-encoder approach to skip the EM-like step in traditional methods for estimating poses. As opposed to an EM approach, the number of encoder parameters is fixed and we do not require explicit estimation of pose distribution for each measurement. Therefore, the number of variables to be estimated in our method does not grow with the number of measurements.

\section{Background and Current Methods}

The reconstruction problem of cryo-EM requires estimation of the structure $\M x \in \R^{N\times N \times N}$ from the measurements $\{\M y_1,\ldots, \M y_M\}$. The acquisition of each measurement $\M y_m \in \R^{N\times N}$ can be modelled as
\begin{equation}\label{eq:forwardmodel}
    \M y_m=\underbrace{ C_{\M d_m}*S_{\M t_m}\{P_{\boldsymbol \theta_m}\{\M x\}\}}_{\M H_{\boldsymbol \varphi_m}}+\M n_m,
\end{equation}
where $\M n_m \in \R^{N \times N}$ is the additive noise. The imaging operator  $\M H_{\boldsymbol \varphi_m}$ depends on the imaging parameters $\boldsymbol \varphi_m=(\boldsymbol \theta_m, \boldsymbol t_m, \boldsymbol d_m) \in \R^8$. 
The imaging operator consists of the projection operator $P_{\boldsymbol \theta^m}$ which outputs the tomographic projection of the structure rotated by the Euler angles $\boldsymbol \theta_m= (\theta_{m,1},\theta_{m,1} ,\theta_{m,1})$; 
the shift operator $S_{\M t_m}$, which shifts the projection by $\M t_m=(t_{m,1}, t_{m,2})$ consisting of horizontal and vertical directions, respectively; 
and the convolution operator $C_{\M d_m}$ which corrupts the image by the contrast transfer function (CTF)  with parameters $\M d_m=(d_{m,1},d_{m,2}, \alpha_{m})$ that consists of major defocus, minor defocus, and astigmatism angle, respectively.

As discussed earlier, the imaging parameters, $\boldsymbol \theta_m$ and $\boldsymbol t_m$, are unknown for each measurement. This coupled with the loss of information due to the CTF, and the high level of noise, makes the reconstruction of $\M x$ a challenging problem.\\

\noindent \textbf{Maximum Likelihood.} A naive approach to solving the reconstruction problem would consist in searching for the unknown imaging parameters for each measurement and a global structure which maximizes the likelihood of the measurements. This quest can be described as
\begin{align}
 \M x_{\rm rec}, \Tilde{\boldsymbol \varphi}_1,\ldots, \Tilde{\boldsymbol \varphi}_m &=\arg \max_{\M x, \Tilde{\boldsymbol \varphi}_1,\ldots, \Tilde{\boldsymbol \varphi}_M } \sum_{m=1}^M \log p(\M y_m |\M x, \Tilde{\boldsymbol \varphi}_m), \\
 &=\arg \max_{\M x, \Tilde{\boldsymbol \varphi}_1,\ldots, \Tilde{\boldsymbol \varphi}_M } \sum_{m=1}^M \|\M y_m -\M H_{\Tilde{\boldsymbol \varphi}_m} \M x\|^2,  \label{eq:ml-noise}
\end{align}
where  $p(\M y_m|\M x, \Tilde{\boldsymbol \varphi}_m)$ denotes the likelihood of the measurement $\M y_m$ given the imaging parameter $\Tilde{\boldsymbol \varphi}_m$ and the structure $\M x$. For an independent white Gaussian noise model, it takes the form \eqref{eq:ml-noise}.

However, since the formulation \eqref{eq:ml-noise} is highly non-convex and filled with poor local minima \cite{ullrich_differentiable_2019}, this naive approach would rarely give a reasonable solution.
A brute-force approach to solving the reconstruction problem would follow an iterative procedure where instead of estimating all the parameters at the same time, the structure would be updated using the current estimates of the poses which would be updated in turn using the current estimate of the structure. 
At iteration $k$, this is given by 
\begin{align}
    \Tilde{\boldsymbol \varphi}_{m,k}&=\arg \max_{\Tilde{\boldsymbol \varphi}_{m}} \log p(\M y_m |\M x_k, \Tilde{\boldsymbol \varphi}_m ) \forall m \in [1, \ldots, M],\\
    \M x_{k+1}&=  \arg \max_{\M x} \log p(\M y_m|\M x, \Tilde{\boldsymbol \varphi}_{k,m} )
\end{align}
This approach would not guarantee to find a global optimum either and the quality of the reconstruction would be highly dependent on the initialization. 
Moreover, the high level of noise makes the pose estimation error-prone.\\

\noindent \textbf{Maximum Marginalized Likelihood.} To remedy these issues, current methods use a marginalized likelihood formulation which instead of estimating a single pose for each measurement, effectively weighs the contribution of the poses from the whole search space. This is given by Eq.\eqref{eq:mml-formulation} where $p(\M y_m|\M x)$ denotes the probability of the projection given the structure.
\begin{align}\label{eq:mml-formulation}
    \M x_{\rm rec}&=\arg \max_{\M x } \sum_{m=1}^M \log p(\M y_m|\M x),\\\nonumber
    &=\arg \max_{\M x } \sum_{m=1}^M \log \int p(\M y_m |\M x, \boldsymbol \varphi) p( \boldsymbol \varphi) \,\mathrm{d} \boldsymbol \varphi,
\end{align}
Formulation \eqref{eq:mml-formulation} can be solved in two stages. 
First, a reasonable approximation to the global maximum can be found through SGD or some other method. Next, this \textit{ab-initio} structure $\M x_{0}$ is  used to initialize an iterative EM procedure (also called iterative refinement) which will refine the solution further, at the cost of doing more operations for pose estimation. 

The expectation step of the $k$-th iteration estimates a conditional distribution on the space of poses for each projection given the current estimate of the structure $\M x_k$. This estimate is given by
\begin{align}\label{eq:e-step}
    p(\boldsymbol \varphi|\M x_k, \M y_m )=\frac{p(\M y_m|\M x_k,\boldsymbol \varphi ) p(\boldsymbol \varphi )}{\int_{\boldsymbol \varphi}p(\M y_m|\M x_k,\boldsymbol \varphi ) p(\boldsymbol \varphi ) \,\mathrm{d} \boldsymbol \varphi}.
\end{align}
The maximization step then uses these poses to update the structures by solving
\begin{align}\label{eq:m-step}
\M x_{k+1}=\arg \max_{\M x}\sum_{m=1}^M\mathbb{E}_{p(\boldsymbol \varphi|\M x_k, \M y_m )}[\log p(\M y_m|\M x,\boldsymbol \varphi ) ].
\end{align}
For feasibility, the space of $\boldsymbol \varphi$ is discretized to compute the integrals in \eqref{eq:e-step} and \eqref{eq:m-step}.
This weighted form of pose estimation is less sensitive to the initial reference and yields better quality reconstructions. In most methods, prior knowledge over the structure is used to obtain a modified  MAP formulation.
However, the traditional approaches just described are computationally intensive procedures because estimating poses or conditional distributions over them against an ever changing reference structure scales poorly - the number of variables to estimate grows directly with the size of the data.

\section{Proposed Method}

To solve the scaling problem, we propose a neural network based representation of poses where we consider the unknown poses $\Tilde{\boldsymbol \varphi}_m$ as the output of a tunable function $E_{\gamma}$ for a given input measurement. In this work, we consider the shifts $\M t_m$ and defocus parameters $\M d_m$ known, but in principle they could also be added to $E_{\gamma}$ outputs.\\

For a general error function $R$, we solve \eqref{eq:proposed-loss-R} which becomes \eqref{eq:proposed-loss} (similar to \eqref{eq:ml-noise}) when a Gaussian noise model is assumed
\begin{align}
 \Tilde{\boldsymbol \varphi}_m&=E_{\gamma}(\M y_m) \quad \forall \, m \in [1, \ldots, M]. \\
 \M x_{\rm rec}&=\arg \max_{\M x, \gamma } \sum_{m=1}^M R( y_m, \M H_{E_{\gamma}(\M y_m)} \M x).  \label{eq:proposed-loss-R} \\
 &=\arg \max_{\M x, \gamma } \sum_{m=1}^M   \|\M y_m -\M H_{E_{\gamma}(\M y_m)} \M x\|^2. \label{eq:proposed-loss}
\end{align}

\begin{figure}[htbp]
\centering\includegraphics[width=\textwidth]{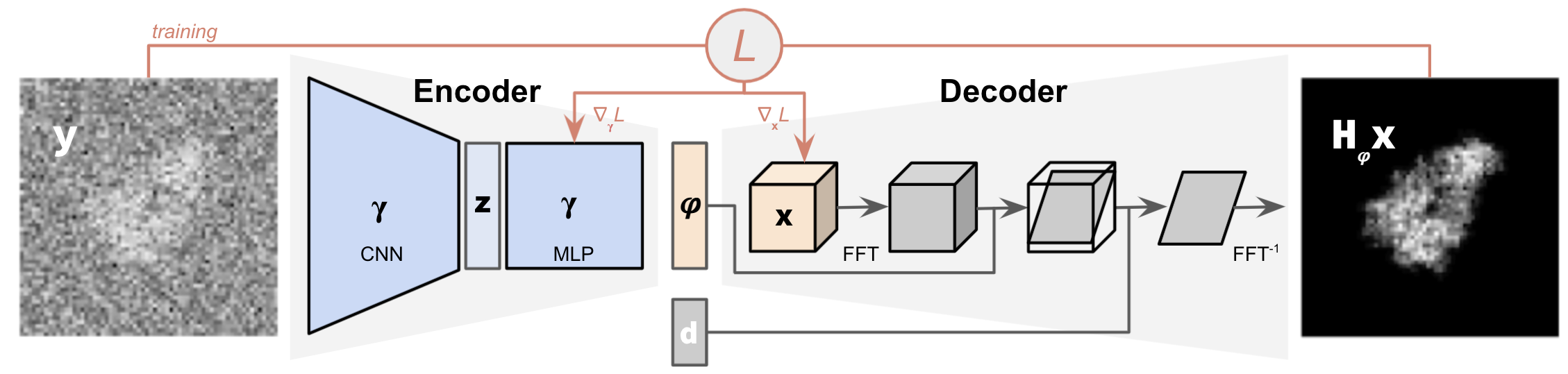}
\caption{\textbf{Auto-encoder architecture}.The input image $\M y$ is encoded into a pose $\boldsymbol \varphi$ by the encoder. The decoder outputs the tomographic projection of the structure oriented at $\varphi$, and then corrupted by the CTF. The image formation is implemented using the Fourier-slice theorem, with the estimated 3D structure $\M x$ Fourier transformed, sliced into a plane normal to $\boldsymbol \varphi$, and corrupted by a known CTF with parameter $\M d$. The parameters of the encoder and decoder are respectively the weights $\boldsymbol \gamma$ of the CNN and MLP and the structure $\M x$. They are optimized during training through minimization of the loss $L$ that measures the dissimilarity between the input image $\M y$ and the reconstructed one $\M H_{\varphi} \M x$.}
\label{fig:architecture}
\end{figure}

We minimize \eqref{eq:proposed-loss-R} using SGD as described in Algorithm \ref{CryoNet-algorithm}. At each iteration, for a batch of measurements $\{\M y_1, \ldots, \M y_B\}$, we get the empirical estimate of the loss function in \eqref{eq:proposed-loss-R} by 
\begin{align}\label{eq:empirical estimate}
    L(\M x, \gamma)=\sum_{b=1}^B R(y_b, \M H_{E_{\gamma}(\M y_b)} \M x).
\end{align}
The structure and  the encoder weights are optimized using the gradients $\nabla_{\M x}L(\M x, \gamma)$ and $\nabla_{\gamma}L(\M x, \gamma)$, respectively. In summary, simultaneously with structure estimation we tune the weights of a neural network so that the imaging parameters maximize the likelihood of the measurements. Our scheme is shown in Figure \ref{fig:architecture}.\\

\begin{algorithm}[ht]
	\caption{Reconstruct cryo-EM data}
	\label{CryoNet-algorithm}
\textbf{Input}: acquired dataset $\{\M y_1,\ldots,\M y_M\}$ ; number of reconstruction iterations, $n_\mathrm{rec}$; size of the batches used for SGD, $B$; optimizer parameters; Error function $R$;\hspace*{\fill} \\
\textbf{Initialization}: $\M x_{\rm rec}$ with a uniform random distribution, and $E_{\gamma}$ with random weights \hspace*{\fill}
\algrenewcommand\algorithmicindent{0.75em}%
\begin{algorithmic}[1]
\For{$n_\mathrm{rec}$}
    \State From acquired dataset, sample a batch$\{\M y^1_{\rm batch},\ldots,\M y_{\rm batch}^B\}$ 
    \State Obtain $\Tilde{\boldsymbol \varphi}_b=E_{\gamma}(\M y_{\rm batch}^b) \forall b \in [1, \ldots, B]$
    \State Compute  $L(\M x, \gamma)=\sum_{b=1}^B R(\M y_{\rm batch}^b,\M H_{\Tilde{\boldsymbol \varphi}_b} \M x_{\rm rec})$
    \State Update $\M x_{\rm rec}$ using $\nabla_{\M x}L(\M x, \gamma)$
    \State update $\gamma$ using $ \nabla_{\gamma}L_(\M x, \gamma)$ \
\EndFor
\end{algorithmic}
\textbf{Output: }$\M x_{\rm rec}, \Tilde{\boldsymbol \varphi}_1, \ldots, \Tilde{\boldsymbol \varphi}_M.$ \hspace*{\fill}
\end{algorithm}

\noindent \textbf{Encoder.} The encoder NN is composed of a convolutional neural network (CNN) followed by a multilayer percepetron (MLP), also known as a fully-connected layer. The CNN extracts shift-invariant features from the measurements, which are then transformed by the MLP into orientation parameters as described below. We use a standard CNN encoder architecture of a \textit{conv2D->conv2D->maxpool} block repeated three times. All convolution layers use $3\times3$ filters, and maxpooling downsamples by a factor of $2$. The number of convolution filters per block is $32, 64, 128$, respectively, and the MLP is composed of two layers, each containing $512$ units/neurons. \\

\noindent \textbf{Differentiable cryo-EM physics model.}
The cryo-EM physics operator $\M H_{\boldsymbol \varphi}$ uses the output of the encoder as the pose parameters ${\boldsymbol \varphi}$.
In order to compute the gradients $\nabla_{\M x}L(\M x, \gamma)$ and $\nabla_{\gamma}L(\M x, \gamma)$, the operator $\M H_{\boldsymbol \varphi}$ needs to be differentiable with respect to the structure as well as the imaging parameters.
We therefore, implement a differentiable cryo-EM physics model which enables the learning of encoder weights using backpropagation. The image formation model is implemented in reciprocal space, in which the the 3D Fourier transform of $\M x$ is computed once per batch. Predicted/Decoded projections are generated by the 2D inverse Fourier transform of a slice in the 3D Fourier volume extracted using the corresponding predicted pose. 
Since the encoder is optimized using backpropagation, the parameterization of the poses directly affects the encoder's performance and thereby, the quality of the reconstruction.
We use the following parameterizations in our experiments:
\begin{itemize}
    \item Euler Angles: For each measurement, the encoder yields a 3-dimensional output which is then fed to the imaging model as Euler angles (commonly used in cryo-EM software). The structure is then rotated using these angles.
    \item Quaternions: The three-dimensional MLP output is transformed into a 4-dimensional vector using a fixed transformation which has the properties of a unit quaternion $\M q= (q_1, q_2, q_3, q_4)$. The imaging model then rotates the structure by $2 \cos^{-1}{q_1}$, around the axis $(q_2, q_3, q_4)$. 
    \item $\mathcal{S}^2\times\mathcal{S}^2$ (s2s2): The MLP outputs two 3-dimensional vectors. These are then orthonormalized to obtain $\M w_1$ and $\M w_2$. A third vector is then computed by the cross product of the first two, $\M w_3 = \M w_1 \times \M w_2$.  These three vectors define a local coordinate system that relates to a $3\times3$ rotation matrix. This matrix is then used to rotate the structure.
    Formally, this case is an algebraic parameterization of the Lie group of 3D rotations SO(3), using two orthonormal vectors $\M w_1$ and $\M w_2$. 
\end{itemize}

NN pose estimation offers many advantages. 
First, instead of searching for individual independent poses $\{\boldsymbol \varphi_m\}$, we estimate a global function that maps the measurements to their respective poses. Since this mapping is global, it indirectly integrates the information from all the projections for each pose estimation. 
Second, since the information of poses has been condensed in the fixed size $\gamma$, the number of variables to estimate does not grow with the size of the data.
Finally, the universal approximation property of the neural networks lets us learn complicated mappings between projections and their poses.\\

\noindent \textbf{Simulation of datasets.} All the experiments presented here used datasets generated from the atomic model 4AKE\footnote{https://www.rcsb.org/structure/4AKE} of \emph{E.coli} adenylate kinase \cite{muller_adenylate_1996}, a small 47 kDa protein. TEM simulator \cite{rullgard_simulation_2011} was used to generate a $128 \times 128 \times 128$ electrostatic potential map with pixel size 0.8 \AA. Images associated with pose $\varphi$ were obtained by rotating the centered map with $\varphi$ and projecting it along the $z$-direction in real space after resampling on the original grid. To account for CTF corruption, the resulting image was Fourier transformed, padded to double length, multiplied by a pre-computed 2D CTF image with given defocus $d$. Gaussian white noise is added \emph{a posteriori}. Datasets are generated by sampling SO(3) uniformally 10,000 times, amounting to 9,000 images for training, and 1,000 for testing. When relevant, the CTF range [0.4 $\upmu$m, 1.2 $\upmu$m] is sampled uniformly. The noise is sampled from the normal distribution and scaled to match the desired signal-to-noise ratio (SNR). Table \ref{table:datasets} lists the datasets generated.\\

\begin{table}[h!]
\centering
 \begin{tabular}{| c || c | c | c |} 
 \hline
 Dataset & SNR (dB) & CTF & size\\ [0.5ex] 
 \hline\hline
 A & - & - & $9,000 \times 128 \times 128$ \\ 
 \hline
 B, C, D, E, F & [10, 5, 0, -5, -10] & - & $9,000 \times 128 \times 128$  \\
 \hline
 G & 0 & [0.4 $\upmu$m, 1.2 $\upmu$m] & $9,000 \times 128 \times 128$  \\ [1ex] 
 \hline
\end{tabular}
\caption{Datasets used in the numerical experiments}
\label{table:datasets}
\end{table}

\noindent \textbf{Training.} For all the described datasets, training was carried with the following parameters: we use the Adam (Adaptive Moment Estimation) \cite{kingma_adam_2017} optimizer with a learning rate of \SI{5e-10}, minibatch size $B = 32$, number of minibatch update steps $n_\mathrm{rec} = 10,000$ which is equivalent to approximately $35.5$ epochs or passes over the whole dataset. The framework is implemented using Tensorflow and runs on an NVIDIA Tesla V100 GPU in around 5 hours for a full training run.

\section{Results}

We tested our method on simulated datasets. We detail below how the methods performed in the absence of noise, in the presence of increasing noise, and in a more realistic setting with noise and CTF corruption.\\

\begin{figure}[htbp]
\centering\includegraphics[width=0.5\textwidth]{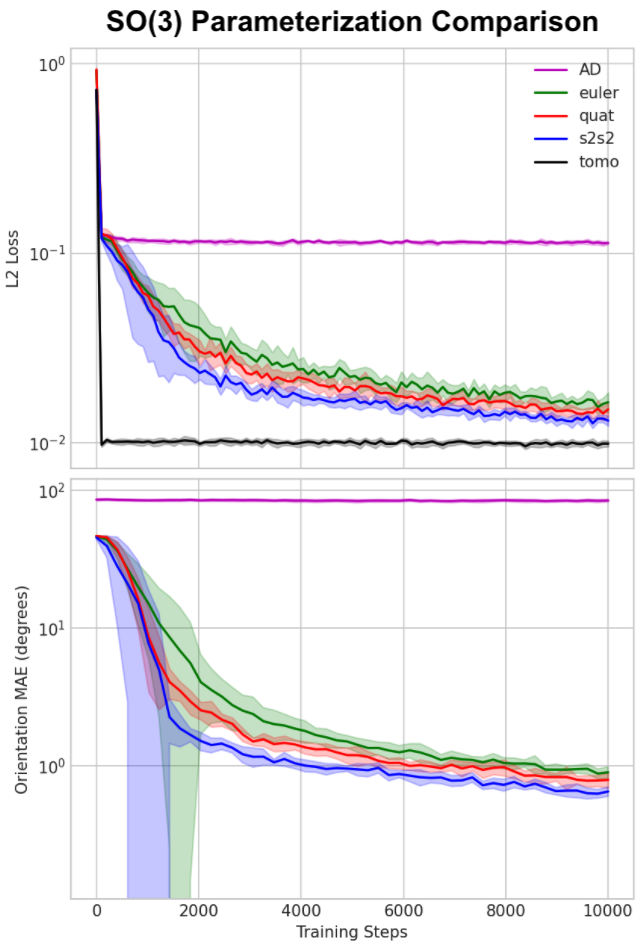}
\caption{\textbf{SO(3) parameterization and convergence}. \textbf{(top)} L2 loss convergence for various representations of SO(3): Euler angles (euler), quaternions (quat) and s2s2. Each corresponds to 10 independent training runs (average and standard deviation). For comparison, the loss curves obtained through automatic differentiation (AD) or with known poses (tomo) are also shown. \textbf{(bottom)} Evolution of the mean absolute error (MAE) over the poses. }
\label{fig:so3parameter}
\end{figure}

\noindent \textbf{SO(3) parameterization.} In the absence of noise (dataset A), training converged in a few thousand steps, as can be seen from the loss curve on Fig.\ref{fig:so3parameter}-top. Following a similar convergence pattern, the MAE between estimated and ground truth poses $\varphi$ decreased drastically in the first few hundred steps (see Fig.\ref{fig:so3parameter}-bottom). Three different approach to SO(3) parameterization have been tried, of which the Euler representation converges the slowest and s2s2 the fastest and ultimately most accurate. On close inspection (data not shown), the main reason for this discrepancy is attributed to parameterization singularities arising from both the Euler and quaternion formulations that could affect the numerical stability of gradients computed around those singularities. For a more in depth proof of those formulations the reader is referred to these publications~\cite{falorsi_explorations_2018,kobayashi_foundations_nodate}. In all the subsequent experiments we adopt s2s2 as the chosen orientation parameterization.\\

\noindent \textbf{Comparison to Automatic differentiation.} To provide points of comparisons with previous work, we considered the following two scenarios. First, we monitored the performance of our approach when poses were known (see $tomo$ curve in Fig.\ref{fig:so3parameter}). In this case, as expected, the network converges almost instantly to what we consider a lower bound for the loss. We also considered the case where the poses and map were solved through stochastic gradient descent, similar to \cite{punjani_cryosparc_2017}. In this case, the loss hardly decreases over a few thousand steps of optimization, mainly due to attempting to solve for orientation and structure simultaneously, which is difficult without careful initialization of the solved variables or alternating the updates between them. Additionally, as mentioned earlier, the number of orientation variables to be solved by AD scales with the dataset size.\\

\begin{figure}[htbp]
\includegraphics[width=\textwidth]{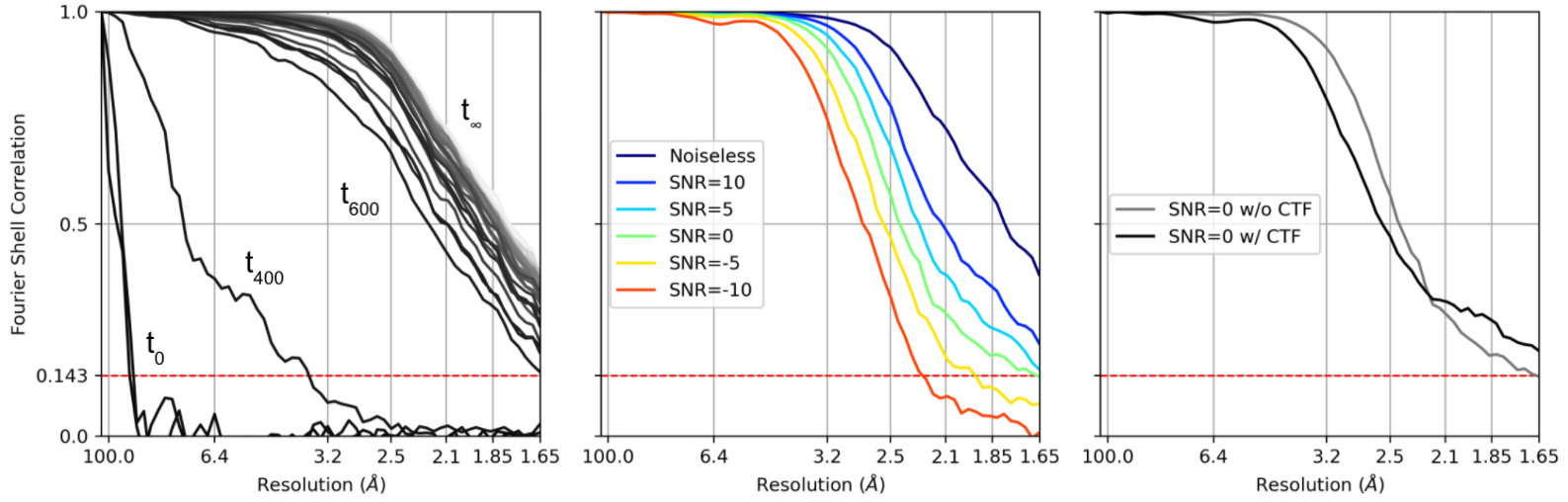}
\caption{\textbf{Fourier-Shell Correlation to ground truth}. The dotted red line indicates the FSC threshold used to estimate the map resolution. \textbf{(left)} FSC curves at regular intervals, from black to white, during optimization in the absence of noise (dataset A). \textbf{(center)} FSC curves for final models at various noise levels (datasets B-F), in the absence of CTF corruption. \textbf{(right)} FSC curves for final models at zero SNR with (dataset D) and without CTF (dataset G) corruption of the dataset.}
\label{fig:fsc}
\end{figure}

\noindent \textbf{Reconstruction quality.} To measure the reconstruction quality of the estimated map $\M x$, we compute the Fourier shell correlation (FSC) every 200 steps between the current and the ground truth map (see Fig.\ref{fig:fsc}-left). In the first few hundred steps, the FSC curve intersects the 0.143 cutoff at resolutions worse than 3.2 \AA~and the Nyquist-Shannon limit of 1.6 \AA~is reached after 600 steps. The FSC keeps improving moderately until it reaches a stable curve.\\

\begin{figure}[htbp]
\centering
\includegraphics[width=0.8\textwidth]{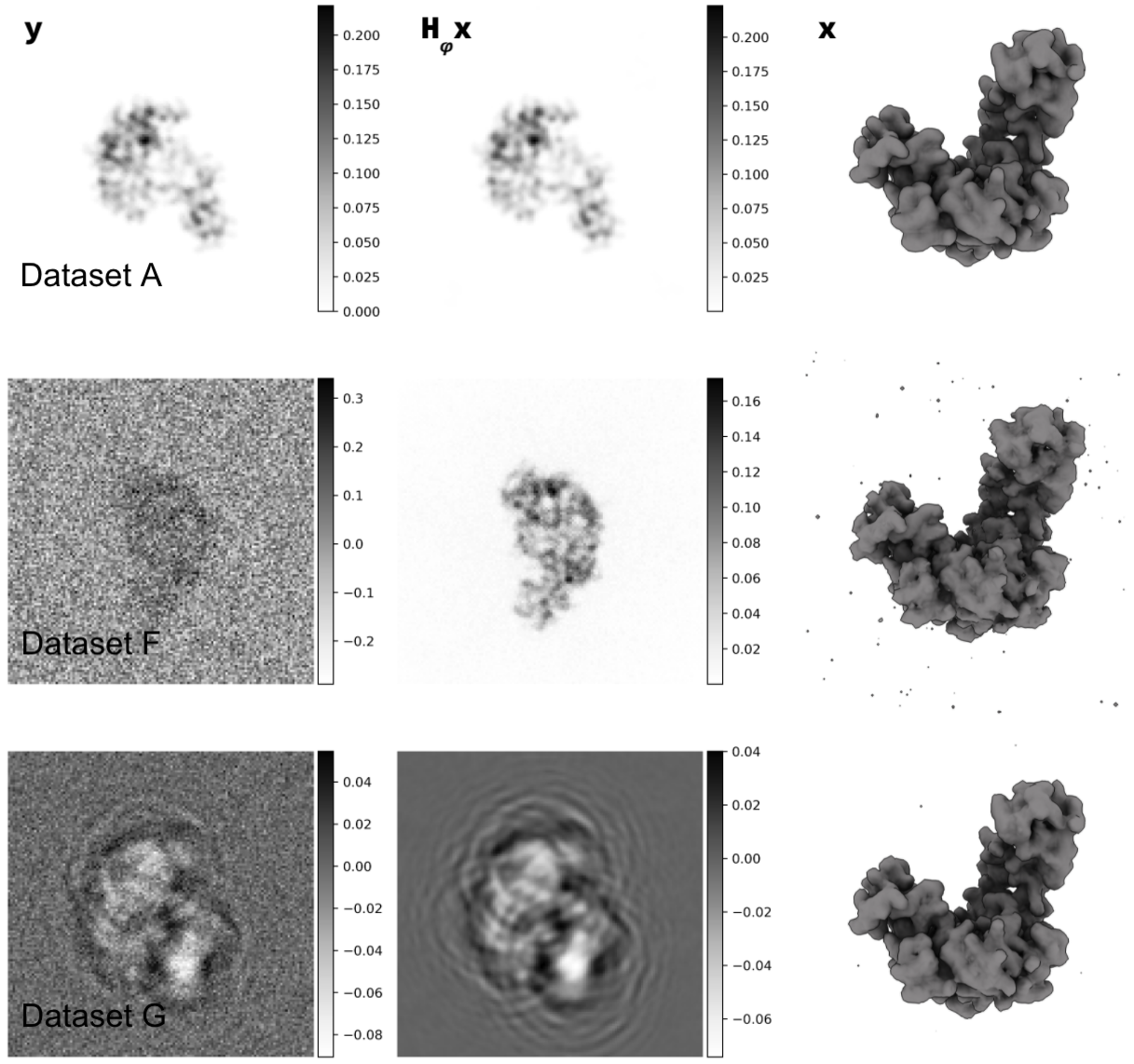}
\caption{\textbf{Reconstructions}. The left column shows samples from their respective datasets (A-top, F-center, G-bottom). The center column shows the reconstructed image after training. The right column shows the reconstructed map at the same contour level.}
\label{fig:recon}
\end{figure}

\noindent \textbf{Effect of noise on reconstruction.} We tested the ability of our method to handle realistic noise typically encountered in cryoEM data. Figure \ref{fig:fsc}-center summarizes our findings. As expected, adding noise is detrimental to the quality of the reconstruction. Yet we see that for the relatively small dataset size that we tried, the Nyquist-Shannon limit is still hit at SNR levels of 0 dB. At more realistic value of -10 dB we see that the effective resolution of the reconstruction is still better than 2.5 \AA. Visual inspection of the reconstructed maps (see Fig.\ref{fig:recon}) is consistent with this observation.\\

\noindent \textbf{Effect of CTF.} Finally, we tested the ability of our method to reconstruct images that were both degraded by added noise (0 dB) and by CTF corruption (dataset G). The FSC curve of the resulting reconstruction is very similar to the one obtained without CTF corruption, both hitting the Nyquist-Shannon limit (see Fig.\ref{fig:fsc}-right). However, the shape differs slightly, with higher correlation in the highest resolution shells and lower correlation in the medium-resolution range.Visual inspection of the resulting map is consistent with this observation (see Fig.\ref{fig:recon}).

\section{Conclusion}

We have presented a new method for solving the pose estimation problem in cryoEM using an autoencoder architecture employing a differentiable forward model decoder. We have shown, on simulated data, that the method was able to reconstruct high-resolution 3D map from simulated data in the presence of noise or CTF corruption and without any prior structural knowledge. We have recently become aware of contemporaneous work using a VAE for estimation of pose and conformation\cite{rosenbaum_inferring_2021}. While this approach also uses NNs to estimate poses, it shares with \cite{zhong_cryodrgn_2021, punjani_3d_2021} the limitation that it relies on a consensus structure initialization, the quality of which might impact the ability of the methods to reliably learn poses and conformations, in particular for highly flexible systems. We hypothetize that the methods presented here would provide a solution to this issue.
Because the computational cost is independent of the number of images, we also envision this method to be the basis of a new pipeline for the large datasets expected to be needed to resolve continuous conformations.

\begin{acks}
We thank Wah Chiu, Khanh Dao Duc, Mark Hunter, TJ Lane, Julien Martel, Nina Miolane, and Gordon Wetzstein for numerous discussions that helped shape this project. This work was supported by the U.S. Department of Energy, under DOE Contract No. DE-AC02-76SF00515. We acknowledge the use of the computational resources at the SLAC Shared Scientific
Data Facility (SDF).
\end{acks}

\bibliographystyle{ACM-Reference-Format}
\bibliography{compSPI}

\end{document}

%% file: Preamble.tex
\usepackage[utf8]{inputenc}

\usepackage{ dsfont,color }
 \usepackage{url}

\usepackage{color}
\usepackage{float}
\newfloat{algorithm}{t}{lop}
\usepackage{amsfonts,amsmath,graphicx,color,epstopdf,amsthm}
\usepackage{float} 
\usepackage{xspace}
\usepackage{url}
\usepackage{booktabs} 
\usepackage{tikz}
\usetikzlibrary{positioning,calc}

\usepackage{graphicx,epsfig}
\usepackage{amsmath,mathrsfs} 
\usepackage{algorithm}
\usepackage[noend]{algpseudocode}
\makeatletter
\def\BState{\State\hskip-\ALG@thistlm}
\makeatother
\usepackage{amsthm}
\usepackage{amsfonts}
\usepackage{mathtools}
\usepackage{multirow}
\usepackage{mathtools} 
\usepackage{epstopdf}
\usepackage{ifthen}
\usepackage{bbm}
\usepackage{setspace}

\newcommand{\R}{ \mathbb{R}}

\def\M#1{{\bf{#1}}}  

\def\R{{\mathbb{R}}}

\makeatletter
\DeclareRobustCommand\onedot{\futurelet\@let@token\@onedot}
\def\@onedot{\ifx\@let@token.\else.\null\fi\xspace}

\makeatother

